\documentclass[aps,preprint,superscriptaddress]{revtex4}

\usepackage{graphicx}
\usepackage{amsmath}

\begin{document}

\title{Skewed distributions as limits of a formal evolutionary process}

\author{F. Sattin}
\email{fabio.sattin@igi.cnr.it}
\affiliation{Consorzio RFX, Corso Stati Uniti 4, 35127 Padova, Italy}

\begin{abstract}
Time series of observables measured from complex systems do often exhibit non-normal statistics, their statistical distributions (PDF's) are not gaussian and often skewed, with roughly exponential tails. Departure from gaussianity is related to the intermittent development of large-scale coherent structures. The existence of these structures is rooted into the nonlinear dynamical equations obeyed by each system, therefore it is expected that some prior knowledge or guessing of these equations is needed if one wishes to infer the corresponding PDF; conversely, the empirical knowledge of the PDF does provide information about the underlying dynamics. In this work we suggest that it is not always necessary. We show how, under some assumptions, a formal evolution equation for the PDF $p(x)$ can be written down, corresponding to the progressive accumulation of measurements of the generic observable $x$. The limiting solution to this equation is computed analytically, and shown to interpolate between some of the most common distributions, Gamma, Beta and Gaussian PDF's. The control parameter is just the ratio between the rms of the fluctuations and the range of allowed values. Thus, no information about the dynamics is required.


\end{abstract}

\maketitle 
Signals produced by noisy, random or chaotic systems do fluctuate around their mean value. While the normal (gaussian) curve is the paradigmatic distribution for these fluctuations, there exist several counterexamples of signals produced from complex systems which exhibit non-normal statistical properties: their statistical distributions (PDF's) are not gaussian, often skewed and with roughly exponential tails. Typical examples are Gamma, log-normal, Weibull distributions \cite{johnson}; the fields where they are encountered span practically all the scientific disciplines (plasma physics, meteorology, financial data, oceanography, biology, ...).
Departure from gaussianity is related to the intermittent development of large-scale coherent structures that break the independence between measurements. Ultimately, the existence of these structures is argued on the basis of the equations governing the dynamics of the system, and in particular by their nonlinearities. In some cases the true equations are not known and replaced by phenomenological models. Thus, a prerequisite for inferring the possible shape of the PDF should be some knowledge or guessing about the system's dynamics \cite{pre05,garcia,garcia2,gus,sandberg,krommes,portelli,carbone}. Conversely, one might expect to employ the empirical knowledge about the PDF in order to infer some information about the unknown underlying dynamics of the system studied. It is interesting to note incidentally that--on the one hand--one and the same kind of distribution may appear in totally different fields but, on the other hand, different experiments measuring the same quantity may yield different PDFs: this is the case, for instance, of laboratory plasma physics where  particle density measurements are mostly interpolated by Gamma distributions, but other PDF's are suggested to be suitable candidates as well \cite{pop4,graves,labit,sattin09a,garcia13,ban}. This might imply that the governing equations must admit several possible classes of solutions. 

In this work we propose a different possibility. We argue that the analytical form of the PDF might be fixed by just few gross constraints which are extracted by the measured data, without invoking any detailed knowledge of the underlying dynamics. To make pictorial this claim we resort to the following paradigmatic case: let us imagine to be measuring some positive-definite quantity, i.e. a (number, mass, ...) density. Suppose as well that the measured quantity is fluctuating and the typical amplitude of its fluctuations (the rms) is comparable to the average value. It is obvious that negative (below the average)  fluctuations are penalized, since no less-than-zero values are permitted, whereas no similar constraint holds for the positive (above the average) ones. Hence, one must expect {\it a fortiori} the PDF to be skewed and non-gaussian: a log-normal PDF, e.g., might be appropriate in this instance \cite{jpcs5}. Of course, dynamical equations do contain the constraint of positiveness but our point is that, in this instance, it is not necessary to invoke them. Another intringuing aspect is that it is not necessary that these contraints do reflect an intrinsic property of the system; rather, they might be imposed by the measuring apparatus as well. Hence--in principle--the same system observed by different observers may produce different PDFs.   

We describe the measurement process as an evolutionary process; the evolution does not occur in real time, but in a fictitious time whose flow parametrizes any increase of information about the system, acquired, e.g., via measurements, within a Bayesian viewpoint, and is therefore a fairly generic feature. The evolving quantity is the PDF, obtained by binning the measured signal, and the quantities of interest are the stable solutions of this equation. 

Let us introduce the model. Any scalar signal $z$ is bounded between a minimum and a maximum value, $(a,b)$. The two extrema are either fixed by physical constraints, or are empirically determined by experiments. For the forthcoming analytical manipulations it is convenient to deal with a PDF whose support is semicompact: either $a \to - \infty$ or $b \to + \infty$. If both of them are finite, then it is necessary to transform the measured variable, say switch to $x = \log [(z-a)/(b-a)]$ which converts the initial interval $(a,b)$ into $(-\infty, 0)$. In most scenarios choosing the opposite range $(0,+ \infty)$ would be a more convenient choice, since measured signals are ordinarily defined as positive numbers. However, the difference is immaterial since amounts to just a change of sign of the observable involved. The present choice has the advantage of being consistent with the existing literature on evolutionary dynamics.  

Let thus $x$ be the measured quantity, and $p_0(x)$ its prior statistical distribution, which may have been obtained via some sets of measurements. Any further investigation provides new information about $x$ and, correspondingly, acts to modify $p_0: p_0 \to p$. Formally we can represent this through a state-transition operator:
\begin{equation}
p(x) = \int dy W(x,y) p_0(y)
\label{eq:w}
\end{equation}
After subtracting $p_0$ from both members we arrive to 
\begin{equation}
\begin{split}
\Delta p \equiv p(x)-p_0(x)  & = \int dy \left(W(x,y)- \delta(x-y) \right) p_0(y) \\
 & \equiv \int dy \, \Theta(x,y) p_0(y)
\end{split}
\label{eq:k}
\end{equation}
Eq (\ref{eq:k}) is an instance of a Master Equation \cite{gardiner}.
The function $\Theta$ accounts for the flow of probability from $y$ to $x$. Generically, it is expected to be system-dependent; however, we will look for universal traits that restrict its possible analytical forms. There are actually two constraints; the first one comes by integrating (\ref{eq:k}) over $x$:
\begin{equation}
\begin{split}
 \int dx \Delta p = 0 \\
& \to \int dx \int dy \; \Theta(x,y) \, p_0(y) = 0  \quad \forall p_0(y)  
\end{split}
\end{equation}
Secondly, the diagonal part of the operator, $\Theta(x,x)$ accounts for a flow of density $p_0$ into itself. Since this is already accounted for by $p_0$ in the lhs of (\ref{eq:k}), this term must be null: $\Theta(x,x) = 0$.

These two constraints allow to guess a plausible form for $\Theta$. First of all we factorize it as: $\Theta(x,y) \equiv T(x-y) \Sigma(x) $. The first term $T$ quantifies the probability flux along distance $x-y$, regardless of the absolute value of the initial and final point. We must account for inhomogeneity as well, and this is brought by  $\Sigma$, that quantifies the propensity for each point $x$ to act as final target. Let us start guessing an analytical expression for $\Sigma$. 
The propensity may be estimated {\it a posteriori} on the basis of the past history of the system, which is quantified by $p_0$ itself. Thus, we may guess that $\Sigma(x) = \Sigma(p_0(x))$, and we choose the simplest proxy: $\Sigma(x) =\Sigma_0 \times p_0(x)$. The function $T$ comes accordingly: the simplest choice compliant with the above requirements is
\begin{equation}
T(x-y) = T_0 \times (x - y) 
\label{eq:t}
\end{equation}
In conclusion, eq. (\ref{eq:k}) rewrites
\begin{equation}
\Delta p = \Delta t (x - <x>) p_0(x) , \quad \Delta t = T_0 \Sigma_0
\label{eq:dp}
\end{equation}
Notice that $\Delta t$ does not depend from $x$ nor $p$; actually, it quantifies the degree of difference between the two successive estimates $p_0, p$: by assuming that each set of measurements increases only marginally the information about $p$, {\it i.e.} $\Delta p \to $, we require therefore $\Delta t \to 0$, and $\Delta t$ may thus be attributed the meaning of an infinitesimal increment of a variable $t$ that parametrizes the state of knowledge about $x$. Eq. (\ref{eq:dp}) takes formally the expression of a finite-difference equation, which may further be turned into a differential equation:
\begin{equation}
{\Delta p \over \Delta t} \approx {\partial p \over \partial t} = (x - <x>) p(x,t)
\label{eq:replicator}
\end{equation}
Equation (\ref{eq:replicator}) is a non-linear integro-differential equation, often encountered in biosciences and game theory, where is known as "replicator equation" \cite{taylor}. In those contexts, $t$ is the true time whereas in our case it is a fictitious time, parameterizing the flow of information.
Several different choices for $\Theta$ other than $T,\Sigma$ above are obviously possible in principle. However, they are penalized in terms of complexity criteria: Eq. (\ref{eq:replicator}) is not the only possible evolution equation but is the simplest one.    

Let us discuss some features of (\ref{eq:replicator}). First of all, the support of $p$ does not vary throughout time: if $p(x) = 0$ at some time for some $x$ then $p(x)$ will be zero at all times. It is possible to generalize Eq. (\ref{eq:replicator}) by including the possibility for $p$ to spread over larger and larger intervals of $x$ (see \cite{smerlak}) but for the moment we will not discuss this option and will return back to it only in the final part of this work. Secondly, Eq. (\ref{eq:replicator}) admits a Dirac-delta stationary solution: $ p_s(x) = \delta (x - x_0) $. Within our picture, this solution corresponds to the case where enough information has been collected to assign a univocal value to the observable $x$: $ t \to \infty$. In actual situations, this never happens. However, in our analysis it will not be necessary taking explicitly this limit: we will show that $p(x,t)$, regardless of initial conditions, approaches quickly  a limiting function that remains unaltered in shape as time flows. 

\begin{figure}
\includegraphics[width=70mm]{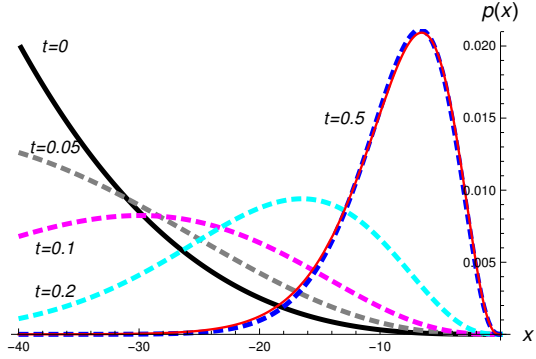}
\caption{An example of distribution evolving from an initial state (solid black curve labelled "$t = 0$") according to Eq. (\ref{eq:replicator}). The final state, at $t = 0.5$ is fitted with a flipped gamma $f(x) = C_0 \exp(C_1 x) (-x)^{C_2}$ (solid red curve). For numerical reasons, obviously, the $x$ range is truncated to finite values, here $(-40,0)$.}
\label{fig:uno}
\end{figure}

Smerlak and Youssef, through a lengthy analysis worked out analytically the solution to Eq. (\ref{eq:replicator}) for large times. We do not replicate here their study, rather we follow the other way round by plugging their result into Eq. (\ref{eq:replicator}) and showing that it is a valid solution {\it at all times}.   
The candidate solution takes the form of a flipped Gamma function:
\begin{equation}
p(x,t)  = {C(t)^{\gamma+1} \over \Gamma(\gamma+1)} \exp \left[C(t) x \right] (- x)^\gamma
\label{eq:gamma}
\end{equation}
The parameter $C$ is conveniently expressed in terms of the average value $\mu = <x>$:
\begin{equation}
C(t) = -{\gamma +1 \over  \mu(t)}
\label{eq:c}
\end{equation}
In Eq. (\ref{eq:c}) we have anticipated that $C$ is function of time.
By taking the time derivative of the logarithm of (\ref{eq:gamma}) we get
\begin{equation}
{\partial p /\partial t \over p} = {-( 1 + \gamma)  \mu \over  \mu^2} \mu'(t)
\end{equation}
which reduces to (\ref{eq:replicator}) when
\begin{equation}
\begin{split}
{(1 +\gamma) \over  \mu^2} \mu'(t)  & = 1 \to \\
 & \mu(t) =  { (1 + \gamma) \mu(0) \over 1 + \gamma - t  \mu(0)}
\end{split}
\label{eq:mu}
\end{equation}
It is easy to check that: (i) $\mu(t) \to 0 $ as $ t \to \infty$, hence $p$ converges towards $p_s = \delta(x)$ and $p$ is self-similar: it does not change shape as time grows; (ii) the rate of convergence slows down and is very small as time increases: $\mu'(t) \propto t^{-2}, t \to \infty$. On the other hand, $\mu$ varies rapidly in the first stages, which suggests that--regardless of the starting point $p_0$--$p$ approaches quickly its limiting value (\ref{eq:gamma}). The numerical integration of Eq. (\ref{eq:replicator}) confirms the validity of this guess (see fig. \ref{fig:uno}). Notice that $p$ of Eq. (\ref{eq:gamma}) is a valid solution at all times, but is also a limiting stable solution: if one starts from a different curve, the solution evolves eventually approaching Eq. (\ref{eq:gamma}).

\begin{figure}
\includegraphics[width=70mm]{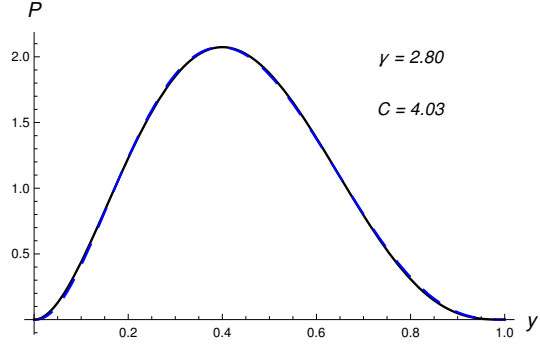}
\caption{The black solid curve is the Beta distribution (\ref{eq:beta}) with $p = 3, q = 4$; the blue dashed curve is the best-fit using Eq. (\ref{eq:pz}). The two curves do overlap perfectly. }
\label{fig:tre}
\end{figure}

Ultimately, we have to revert to the physical variable $z$ by replacing into (\ref{eq:gamma}) $x$ with:
$ x \to \log \left[(z-a)/(b-a)\right]$. The result is 
\begin{equation}
p(z) \propto \left[-\log(y)\right]^\gamma y^{C-1} , \quad y \equiv {z-a \over b - a}
\label{eq:pz}
\end{equation}
This expression is not particularly illuminating, but its limiting forms are revealing. The relevant figures here are the ratios $\sigma/a,b$, where $\sigma$ is a measure of the typical amplitude of the signal: say the average value, or the mean fluctuation around the average. We will consider both $a,b$ to be non-zero in order not to bother with sub-cases. (I) Let us keep $\sigma/b$ finite and take $ |\sigma/a| << 1 $. This obviously corresponds practically to reverting back to the original variable $x$: indeed, $x = \log [(z-a)/(b-a)] \approx \log (1-z/a) \approx z/(-a)$. Hence, $p(z)$ turns into a Gamma function like Eq. (\ref{eq:gamma}). (ii) A second limiting case involves both $a,b$ to be very large: $ |\sigma/a| , |\sigma/b| << 1 $. The function $p(z)$ has its maximum in 
$z_M: \log \left[(z_M - a)/(b-a)\right] = \gamma /(C-1)$.
If we expand $\log p$ around this point we get 
\begin{equation}
\log p/p(z_M) \approx - \left(z - z_M \over b-a \right)^2 + O\left(z - z_M \over b-a \right)^3 
\end{equation}
Thus now, $\sigma = \sqrt{<(z- z_M)^2>}$.
As long as $|\sigma $ remains much smaller than $b-a$, hence, the lowest-order term dominates and $p$ reduces to a gaussian PDF.   
(iii) Finally, let us consider the case when both $\sigma/a$ and $\sigma/b$ are finite. A fairly versatile family of functions, often employed to model distributions within finite domains, is the Beta distribution \cite{johnson}. With our definitions it reads:
\begin{equation}
B = {\Gamma(p + q) \over \Gamma(p) \Gamma(q) } y^{p-1} (1-y)^{q-1}
\label{eq:beta}
\end{equation}
In Fig. (\ref{fig:tre}) we show that (\ref{eq:pz}) indeed does fit remarkably well Eq. (\ref{eq:beta}). This is not unexpected, since, for $ y \approx 1$, we can Taylor expand $- \ln(y) \approx (1-y) + O(1-y)^2$ and there is an explicit correspondence between Eq. (\ref{eq:pz}) and (\ref{eq:beta}): the Beta PDF can be regarded as the lowest-order series expansion of (\ref{eq:pz}). In conclusion, we may claim that even this family of distributions is accounted for within our model.    

\begin{figure}
\includegraphics[width=70mm]{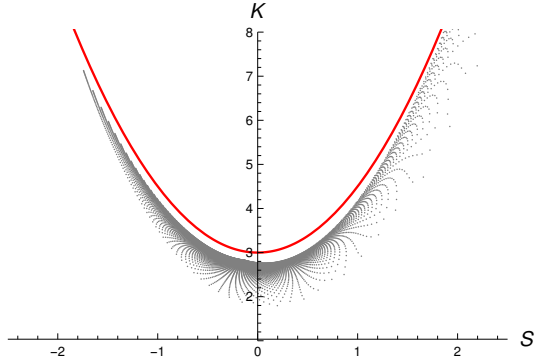}
\caption{The solid red curve is Eq. (\ref{eq:sk}); the gray dots are computed from Eq. (\ref{eq:pz}).}
\label{fig:due}
\end{figure}

Rather than inspecting the whole PDF, an often employed strategy when dealing with experimental data is to study the mutual relations between the lowest-order statistical moments, in particular the third (skewness $S$) and fourth (kurtosis $K$) one.  Low-order statistical moments are robust quantities to compute from raw data; furthermore, several PDFs feature characteristic correlations between them. For the Gamma distribution (\ref{eq:gamma}), in particular, it is easy to show that
\begin{equation}
K = {3 \over 2} \, S^2 + 3 
\label{eq:sk}
\end{equation}
(The gaussian PDF is a special case of this law, with $S \equiv 0, K \equiv 3$).
It is a well established empirical result that data from very diverse fields do obey a law like (\ref{eq:sk}): $ K = A S^2 + B$,  
with $A,B$ close to the values $3/2, 3$, respectively. This scaling has been studied extensively in laboratory plasma physics \cite{labit,sattin09a,ban,sattin09b,yan,kube16}, but also in oceanography \cite{sura}, meteorology \cite{schlop}, seismology and financial time series \cite{cristelli,adcock}. In Fig. (\ref{fig:due}) we plot the $(S,K)$ curve from Eq. (\ref{eq:sk}) together with a large sample of $(S,K)$ couples computed from (\ref{eq:pz}) by varying $\gamma$ and $C$. Expression (\ref{eq:pz}) does not allows for a unique relation between $S$ and $K$, rather a cloud of points is generated, which is bounded from above by the Gamma limiting solution (\ref{eq:sk}). Visual comparison of Fig. (\ref{fig:due}) with the similar ones from experiments shows that that there is a good overlap between the two data sets, but for one aspect: Eq. (\ref{eq:pz}) does not appear to include solutions featuring $K > 3/2 S^2 + 3$, whereas part of the experimental data do lie in this region. In terms of PDFs this implies that datasets do exist whose histograms are not interpolated by none of the distributions studied above. This is, at this stage, a shortcoming of the model but we do not think it is a critical one. One possible way of addressing this issue is by acknowledging that we have so far dealt with immutable ranges $(a,b)$. This is justifed as long $a,b$ are fixed {\it a fortiori} by the physics of the system or by the measuring apparatus. In several situations, on the other hand, $a,b$ are only defined {\it a posteriori} as the extrema of the signal measured so far, hence they are susceptible of varying just like $p$ does. Smerlak and Youssef \cite{smerlak} argue that this effect can be accounted for by adding an effective diffusive term in eq. (\ref{eq:replicator})--converting it into a nonlinear Fokker-Planck equation with reaction terms. Their numerical studies of this modified equation show that--transiently--its solutions do move in the $(S,K)$ plane above the curve $3/2 S^2 + 3$.            

In conclusion, we are suggesting that the statistical distributions encountered in the analysis of experimental data (with the exception of power-law ones) may arise generically enforced by few natural constraints. Our whole discussion relies basically on just the three ansatzs defining the functions $\Theta$, $\Sigma$, $T$. The appealing consequence of our hypothesis is that we are able to reduce ourselves to just one basic solution, Eq. (\ref{eq:pz}). The best known and most common distributions do arise just as several different cases of this solution. The relevant parameters for interpolating between the one and the other limits are, roughly, the ratios between the typical measured values (say, the rms) and the maximum measurable ones, with the Gaussian PDF appearing when both ratios go to zero. Thus we able to claim that the universality of PDFs across several different systems is possible, while at the same time providing a rationale for one kind of signals being modelled by different curves in different experiments.

\begin{acknowledgments} 
The author wishes to thank S. Cappello, D. Escande, N. Vianello, M. Agostini, M. Zuin and I. Predebon for useful discussions
\end{acknowledgments}


\begin{thebibliography}{24}

\bibitem{johnson} N.L. Johnson, S. Kotz, N. Balakrishnan, {\it Continuous Univariate Distributions} (John Wiley \& Sons, New York,
1995), 2nd ed
\bibitem{pre05} F. Sattin, N. Vianello, {\it Phys. Rev. E} {\bf 72} (2005) 016407
\bibitem{garcia} O.E. Garcia, {\it Phys. Rev. Lett.} {\bf 108} (2012) 265001
\bibitem{garcia2} O.E. Garcia, R. Kube, A. Theordosen, H.L. P\'ecseli, {\it Phys. Plasmas} {\bf 23} (2016) 052308
\bibitem{gus} D. Guszejnov, N. Laz\'anyi, A. Bencze, S. Zoletnik, {\it Phys. Plasmas} {\bf 20} (2013) 112305
\bibitem{krommes} J.A. Krommes, {\it Phys. Plasmas} {\bf 15} (2008) 030703
\bibitem{portelli} B. Portelli, P.C.W. Holdsworth, J.-F. Pinton, {\it Phys. Rev. Lett.} {\bf 90} (2003) 104501
\bibitem{carbone} V. Carbone, {\it et al}, {\it Europhys. Lett.} {\bf 58} (2002) 349
\bibitem{sandberg} I. Sandberg, S. Benkadda, X. Garbet, G. Ropokis, K. Hizadinis, D. del-Castillo-Negrete, {\it Phys. Rev. Lett.} {\bf 103} (2009) 165001
\bibitem{pop4} F. Sattin, N. Vianello, M. Valisa, {\it Phys. Plasmas} {\bf 11} (2004) 5032
\bibitem{graves} J.P. Graves, J. Horacek, R.A. Pitts, K.I. Hopcraft, {\it Plasma Phys. Control. Fusion} {\bf 47} (2005) L1
\bibitem{labit} B. Labit, I. Furno, A. Fasoli, A. Diallo, S.H. M\"uller, G. Plyushchev, M. Podest\'a, F.M. Poli, {\it Phys. Rev. Lett.} {\bf 98} (2007) 255002
\bibitem{sattin09a} F. Sattin, M. Agostini, R. Cavazzana, P. Scarin, J.L. Terry, {\it Plasma Phys. Control. Fusion} {\bf 51} (2009) 055013
\bibitem{garcia13} O.E. Garcia, S.M. Fritzner, R. Kube, I. Cziegler, B. LaBombard, J.L. Terry, {\it Phys. Plasmas} {\bf 20} (2013) 055901
\bibitem{ban} S. Banerjee, {\it et al}, {\it Phys. Plasmas} {\bf 21} (2014) 072311
\bibitem{jpcs5} F. Sattin, N. Vianello, M. Valisa, V. Antoni, G. Serianni, {\it Journal of Physics: Conf. Series} {\bf 7} (2005) 247
\bibitem{gardiner} C.W. Gardiner, {\it Handbook of Stochastic Methods} (Springer, 2009)
\bibitem{taylor} P.D. Taylor and L.B. Jonker, {\it Math. Biosci.} {\bf 40} (1978) 145
\bibitem{smerlak} M. Smerlak and A. Youssef, {\it J. Theor. Biol.} {\bf 416} (2017) 68
\bibitem{sattin09b} F. Sattin, M. Agostini, R. Cavazzana, G. Serianni, P. Scarin, N. Vianello, {\it Phys. Scr.} {\bf 79} (2009) 045006
\bibitem{yan} N. Yan {\it et al}, {\it Plasma Phys. Control. Fusion} {\bf 55} (2013) 115007
\bibitem{kube16} R. Kube, A. Theordosen, O.E. Garcia, B. LaBombard, J.L. Terry, {\it Plasma Phys. Control. Fusion} {\bf 58} (2016) 054001
\bibitem{sura} P. Sura and P.D. Sardeshmukh, {\it J. Phys. Oceanogr. } {\bf 38} (2008) 638
\bibitem{schlop} T.P. Schlopflocher and P.J. Sullivan, {\it Boundary-Layer Meteor.} {\bf 115} (2005) 341
\bibitem{cristelli} M. Cristelli, A. Zaccaria, L. Pietronero, {\it Phys. Rev. E} {\bf 85} (2012) 066108
\bibitem{adcock} C. Adcock, M. Eling, N. Loperfido, {\it Eur. J. Finance} {\bf 21} (2015) 1253   

\end{thebibliography}
\end{document}